\documentstyle[12pt,aaspp4]{article}

\lefthead{K.~MIGHELL ET AL.:
LMC GLOBULAR CLUSTER HODGE 11}
\righthead{K.~MIGHELL ET AL.:
LMC GLOBULAR CLUSTER HODGE 11}
\slugcomment{
{Accepted by the Astronomical Journal on 5 March 1996}
}

\begin{document}

\def\et{\hbox{\em{et~al.\ }}}
\def\vs{\hbox{vs }}
\def\bmv{\hbox{$B-V$}}
\def\ebmv{\hbox{E$(\bmv)$}}
\def\feh{\hbox{[Fe/H]}}

\title{
WFPC2 OBSERVATIONS OF STAR CLUSTERS IN THE MAGELLANIC CLOUDS:
I. THE LMC GLOBULAR CLUSTER HODGE 11\altaffilmark{1}
}

\author{
\sc
Kenneth J. Mighell
and
R. Michael Rich\altaffilmark{2}
}
\affil{
Dept.\ of Astronomy,
Columbia University
538 W.\ 120th Street,
New York, NY~~10027\\[-1mm]
Electronic mail:
mighell$@$figaro.phys.columbia.edu,
rmr$@$carmen.phys.columbia.edu
}

\author{
\sc
Michael Shara
and
S. Michael Fall
}
\affil{
Space Telescope Science Institute,
3700 San Martin Drive,
Baltimore, MD~~21218\\[-1mm]
Electronic mail:
mshara$@$stsci.edu,
fall$@$stsci.edu
}

\altaffiltext{1}{
Based on observations made with the NASA/ESA
{\em{Hubble Space Telescope}},
obtained at the Space Telescope
Science Institute,
which is operated by the Association of
Universities for Research in Astronomy, Inc.\ under NASA
contract NAS5-26555.
}
\altaffiltext{2}{
Alfred P. Sloan Foundation Fellow.
}

\begin{abstract}

We present our analysis of
{\sl{Hubble Space Telescope}} Wide Field Planetary Camera 2
observations
in
F555W (broadband $V$)
and
F450W (broadband $B$)
of the globular cluster Hodge 11 in the Large Magellanic Cloud galaxy.
The resulting $V$ vs.\ $\bmv$ color-magnitude diagram reaches
$2.4$ mag below the
main-sequence turnoff
(which is at
$V_{\rm{TO}}=22.65\pm0.10$ mag or $M_V^{\rm TO}=4.00\pm0.16$ mag).
Comparing the fiducial sequence of Hodge 11
with that of the Galactic globular cluster M92,
we conclude that, within the accuracy of our photometry,
the age of Hodge 11 is identical to that of M92
with a relative age-difference uncertainty ranging from 10\% to 21\%.
Provided that Hodge 11 has always been a part
of the Large Magellanic Cloud and was
not stripped from the halo of the Milky Way or
absorbed from a cannibalized dwarf spheroidal galaxy,
then the oldest stars in the
Large Magellanic Clouds and the Milky Way appear to have the same age.
\end{abstract}

\newpage
\section{INTRODUCTION}
The oldest stars for which reliable ages can be determined are found in
globular clusters.  Of the various methods used to infer ages in these
systems, the absolute magnitude of the main-sequence turnoff is considered
to be the most secure measurement.
The main-sequence turnoff point for old stars occurs at
$M_V\approx +4.0$ mag
which makes it difficult to observe the old turnoff population
in relatively nearby galaxies such as Andromeda (M31)
even with the refurbished
{\em{Hubble Space Telescope}}.
The ages of the oldest stars in a cluster of galaxies
can be used to establish the chronology of galaxy formation
within that cluster.
The age-spread of the oldest stars provides
an important cosmological probe for the investigation
of synchronized galaxy formation.
Current technology, unfortunately, allows us to conduct this
experiment only with the Milky Way and its relatively nearby companions.

Efforts have been made to infer the ages of star clusters in the
Magellanic Clouds from their integrated photometry.
The Large Magellanic Cloud (LMC) star cluster Hodge 11 (Hodge 1960)
was classified by
Searle, Wilkinson, \& Bagnuolo (1980) as being SWB class VII
and suggested that Hodge 11 is similar to
the old metal-poor Galactic halo globular clusters.
This was confirmed by
Elson \& Fall (1988)
and
Girardi \et (1995) whose new UBV
cluster photometry places Hodge 11 clearly among
the oldest Galactic globular clusters.
The LMC globular cluster NGC 2257 was recently thought to be similar to
Hodge 11, however Testa \et (1995) has determined that
NGC 2257 is 2--3 Gyr younger than the oldest Galactic globular clusters.
Therefore, Hodge 11 is especially interesting because it might be the
oldest globular cluster in the Magellanic Clouds.

This paper presents the first results of our
{\em{Hubble Space Telescope}}
snapshot observation program of star clusters in the Magellanic
Clouds using the Wide Field Planetary Camera 2 instrument.
Our sample of star clusters was chosen to cover the full age range
available in the Clouds and
we have surveyed 46 star clusters using $\sim$15 hours of spacecraft time.
While the principal aim of our observational program was to investigate
the global properties of star clusters in the Magellanic Clouds,
we now turn to our results on the Large Magellanic Cloud
globular cluster Hodge 11.

\section{OBSERVATIONS AND PHOTOMETRIC REDUCTIONS}

The LMC globular cluster Hodge 11
was observed with the
{\sl{Hubble Space Telescope}} Wide Field Planetary Camera 2 (WFPC2)
on 1994 February 1 through the
F450W (broadband $B$)
and F555W
(broadband $V$)
filters.
The WFPC2 has four internal cameras --- the Planetary Camera (PC) with
a focal ratio of $f/28.3$ and three Wide Field (WF) Cameras at $f/12.9$
(Holtzman \et 1995a).
Each camera images onto
a Loral $800 \times 800$ CCD which gives a plate scale of
0\farcs046 pixel$^{-1}$ for the PC camera and
0\farcs10 pixel$^{-1}$ for the three WF cameras.
The WFPC2 PC1 aperture
(Burrows 1994)
was centered on the target position of
$\alpha = 06^{\rm h}\ 14^{\rm m}\ 23^{\rm s}$
and
$\delta = -69\arcdeg\ 50\arcmin\ 50\arcsec$
(J2000.0)
and
two high-gain observations were obtained:
600-s in F450W
and
300-s in F555W.
The two datasets
(F450W: U26M0F01T;
 F555W: U26M0F02T)
were recalibrated using
the {\tt{calwp2}} task in the
{\tt{stsdas.hst\_calib.wfpc}} package
({\tt{IRAF V2.10.3BETA}} and {\tt{STSDAS Version 10.3.2}})
and the calibrated reference files
are given in Table \ref{tableone}.
All pixels that were flagged ``bad'' in the data quality files
of the original observations
were replaced by linearly interpolated values of nearby ``good'' pixels
by using the
{\tt{wfixup}} task.

\placetable{tableone}

Unsharp masks of the Hodge 11 observations were made
using the LPD (low-pass difference) digital filter
which was designed by Mighell
to optimize the detection of faint stars in {\sl HST} WF/PC and WFPC2
images (Appendix A of Mighell \& Rich 1995, and references therein).
The two unsharp mask images were
then added together to make a master unsharp mask image for each WFPC2 CCD.
A simple peak detector algorithm was then used on the master unsharp
images to create a
list of point source candidates
with coordinates $60 \leq x \leq 790$ and $60 \leq y \leq 790$
on each CCD.
This allowed us to use almost the entire field-of-view of each WFPC2 camera
while avoiding edge-effects in the outer regions.

We analyzed the data with
the {\tt{CCDCAP}} digital circular aperture
photometry code recently developed by Mighell to analyze
{\sl{HST}} WF/PC and WFPC2
observations (Mighell \& Rich 1995, 1996, and Rich \& Mighell 1995).
A fixed aperture with a radius of 1.8 pixels was used for all stars
on the PC and WF CCDs.
The local background level was determined from a robust estimate
of the mean intensity value of all pixels between 1.8 and 4.8 pixels
from the center of the circular
stellar aperture.
Point source candidates were rejected if either
one of two criteria was
satisfied:
(1) the measured signal-to-noise ratio $SNR<5$;
or (2) the center of the aperture (which we allowed to move in order to
maximize the $SNR$) changed by more than 2 pixels
from its detected position on the master unsharp mask.
These criteria allowed us to automatically eliminate
most of the photon noise spikes due to the background
sky and diffraction spikes.
For comparison, we also reduced the data using
the {\tt{DAOPHOT}} program (Stetson 1987) and found that
photometric scatter was significantly
larger in the {\tt{DAOPHOT}} photometry
than in the {\tt{CCDCAP}} photometry.

These observations were obtained when the
WFPC2 CCDs operated at a temperature
-76 $^\circ$C.
At this temperature,
the number of ``hot'' pixels on a WFPC2 CCD
would grow at a rate of several thousand pixels per month per chip
(Holtzman \et 1995a).
There was a small but statistically significant position shift
for stars between the F450W and F555W frames.
Hot pixels and other CCD defects did not exhibit this position shift.
We took advantage of this fact to reject all point-source candidates
with a position shift between the two frames
that was not within five standard deviations of the mean shift on each WFPC2
CCD.  This procedure allowed us to easily remove
almost all of the hot pixels and other CCD defects.

Observed WFPC2 point spread functions (PSFs)
vary significantly
with wavelength, field position, and time (Holtzman \et 1995a).
There were not enough bright isolated stars in our Hodge 11 observations
to allow us to adequately measure the variation of the point spread function
across the WFPC2 CCDs using the observations themselves.
We have determined aperture corrections
($\Delta m_r$)
based on measurements of artificial point spread functions
synthesized by the
{\tt{Tiny Tim Version 4.0b}}
software package (Krist 1993, 1994).
We created a catalog of 289 synthetic M-giant point spread functions
in a $17 \times 17$ square grid
for each
filter (F450W and F555W)
and CCD (PC1, WF2, WF3, and WF4).
The spatial resolution of one synthetic PSF
every 50 pixels in $x$ and $y$ allowed us to determine aperture corrections
for any star in the entire WFPC2 field-of-view to within a worst-case
spatial resolution of 35 pixels.
The average aperture corrections
($\overline{\Delta m_r}$)
are listed in Table \ref{tabletwo}.
\placetable{tabletwo}

The WFPC2 point spread functions can vary with time due to
spacecraft jitter during exposures and small focus
changes caused by the {\sl{HST}} expanding and contracting (``breathing'')
once every orbit.
These temporal variations of WFPC2 PSFs
can cause small, but significant, systematic
offsets in the photometric zeropoints when small apertures are used.
Fortunately,
these systematic offsets can be easily calibrated away by
simply measuring
bright isolated stars on each CCD twice: once with the small aperture
and again with a larger aperture.
The robust mean magnitude difference
between the large and small apertures
is then
zero-order aperture correction
($\Delta m_0$)
for that particular CCD and filter combination
(see Table \ref{tabletwo}).
On the PC1 CCD, we used a large aperture with a radius of 5.0 pixels and
the background was determined using an annulus of
$5.0\leq r_{\rm sky}\leq13.0$ pixels.
On the WF CCDs, we used a large aperture with a radius of 3.0 pixels and
the background was determined using an annulus of
$3.0\leq r_{\rm sky}\leq6.0$ pixels.
These large apertures contain about 86\% of the total
flux from a star.

We use the standard WFPC2 color system
(Holtzman \et 1995b)
which is defined using apertures 1\arcsec\ in diameter that
contain about 90 percent of the total flux from a star.
A final aperture correction
($\Delta m_\infty$)
was added to the instrumental magnitudes
to go from an infinite aperture to an aperture 1\arcsec\ in diameter
(see Table \ref{tabletwo}).
The Charge Transfer Effect
was then removed using a uniform wedge
along the Y-axis of each chip as described in
Holtzman \et (1995b).
Finally,
the instrumental magnitudes, $b$ and $v$,
were transformed to Johnson $B$ and $V$
using the following color equations
\begin{eqnarray}
B &=& b + [0.230\pm0.006](\bmv)\nonumber\\
   && +\ [-0.003\pm0.006](\bmv)^2\nonumber\\
   && +\ [21.175\pm0.002]
\end{eqnarray}
(Table 10 of Holtzman \et 1995b) and
\begin{eqnarray}
V &=& v + [-0.060\pm0.006](\bmv)\nonumber\\
   && +\ [0.033\pm0.002](\bmv)^2\nonumber\\
   && +\ [21.725\pm0.004]
\end{eqnarray}
(Table 7 of Holtzman \et 1995b)
where an instrumental magnitude of zero is defined as one DN/sec at
the high gain state ($\sim$14 e$^-$/DN).

\section{THE COLOR-MAGNITUDE DIAGRAM OF HODGE 11}

The early photographic photometry of Gascoigne (1966) clearly revealed
the unusual nature of Hodge 11: its large number of very blue stars and the
absence of a defined giant branch.
While some electronographic photometry of Hodge 11 was attempted,
little substantive progress occurred until the CCD photometry of
Stryker \et (1984), which clearly identified the blue horizontal
branch and M92-like red giant branch after a careful subtraction
of the substantial contaminating LMC field population.
Walker's (1993) study represents the best ground-based photometry of
Hodge 11 to date but his field-subtracted color-magnitude diagram
only reaches
$V\approx22$ mag which is $\sim$3 mag below the horizontal branch
and $\sim$0.5 mag above the main-sequence turnoff of an old
($\sim$15 Gyr) globular cluster.
We have compared our photometry of giants in Hodge 11 with
the data set of Alistair Walker (1993) which he kindly provided to us.
Using only the WF4 CCD, we find
80 giants in common between the
two data sets.
Figure \ref{figv} shows that
there is a statistically insignificant difference between the $V$ zeropoints
of the two data sets ($0.009\pm0.010$ mag with our $V$ photometry
being $\sim$1\% fainter).
\placefigure{figv}

The $V$ \vs $\bmv$ color-magnitude diagram (CMD) of our observed stellar
field in Hodge 11 reaches $V\approx25.0$ mag and is displayed in
Fig.\ \ref{figi}.
Figure \ref{figi}a shows all the data with
13175 stars (filled circles) and 1855 CCD defects (open circles).
Hodge 11 is surrounded by a field population that is mostly 2--3 Gyr old
with a metallicity of $\feh\approx-0.5$ dex (Walker 1993).
We must remove the contamination by the young LMC field population
before we can proceed to analyze the color-magnitude diagram of Hodge 11.
The  maximum radial distance from the center
of Hodge 11 for our WFPC2 observations
is only 127.4\arcsec.
The core and tidal radii of Hodge 11 are
$r_c=18\arcsec$
and
$r_t=180\arcsec$,
respectively (Mateo 1987).
Our radial coverage of Hodge 11 is thus
$0\leq r\leq7.1r_c$
or
$0\leq r\leq0.71 r_t$
and our observation
is well within the tidal radius of the globular cluster.
We are forced to {\em approximate} the LMC field population by
erroneously assuming
that {\em all} of the stars in the outer regions of our observation
are LMC field stars.
We split our observation into two regions: (1) the cluster region
($r\leq 83.2\arcsec$: 12281 stars --- see Fig.\ \ref{figi}b)
and
(2) the LMC field region
($r > 83.2\arcsec$: 894 stars --- see Fig.\ \ref{figi}c)
which is comparable with the LMC field CMDs of
Stryker \et (1984: Fig.\ 1)
and
Walker (1993: Fig.\ 4).

We used the following procedure to statistically remove the LMC
field population from the cluster region CMD.
Every star in the cluster region CMD
has a $V$ magnitude
with an error $\sigma_V$ and a $\bmv$ color with
an error $\sigma_{(B-V)}$.
For a given star in the cluster region CMD
(Fig.\ \ref{figi}b)
we can count how many stars can be found in that CMD
that have $\bmv$ colors within
MAX$(2\sigma_{(B-V)},0.100)$ mag
and $V$ magnitudes within
MAX$(2\sigma_V,0.200)$ mag.
Let us call that number $N_{\rm H11}$.
We can also count how many
stars can be found in the {\em field} region CMD (Fig.\ \ref{figi}c)
within the same $V$ magnitude range and $\bmv$ color range
that was determined for the star in the {\em cluster} region CMD.
Let us call that number $N_{\rm LMC}$.
The probability, $p$,
that the star in the cluster region CMD
is actually a cluster member of
Hodge 11 can be approximated as follows
\begin{equation}
p \approx 1 - {\rm MIN}
\left(
\frac{\alpha (N_{\rm LMC} +1)}{N_{\rm H11} +1}
,1.0
\right)
\end{equation}
where $\alpha\equiv3.00$ which is the ratio of the area of
the cluster region
(3.58 arcmin$^2$)
to the area of the LMC field region
(1.19 arcmin$^2$).
We see,
by definition,
that
$N_{\rm H11} \geq 1$,
$N_{\rm LMC} \geq 0$,
and
$0\leq p \leq 1$.
Now suppose for a given star in the cluster region CMD that we find
$N_{\rm H11}=78$
and
$N_{\rm LMC}=10$,
then the probability of cluster membership
is $p \approx 0.582$ or 58.2 percent.
We can determine the probable cluster membership of this star
by picking a uniform random number, $0 \leq p^\prime \leq 1$,
and if $p^\prime \leq p$ then
this star is said to probably be a cluster member.
Using a uniform random number generator,
we determined the probable cluster membership for all 12281 stars in
the cluster region CMD field.  Only 9506 stars were retained as
probable cluster members and they are displayed as the cleaned
cluster CMD (see Fig.\ \ref{figi}d).
Since the above CMD cleaning method is probabilistic, this
figure represents only one out of an infinite number of different
realizations of the cleaned Hodge 11 CMD.
The cleaned Hodge 11 CMD diagram (Fig.\ \ref{figi}d)
contains 9506 stars which implies that
about 23\% of the stars in the cluster region CMD
(Fig.\ \ref{figi}b) are LMC field stars.

\placefigure{figi}

The cleaned cluster color-magnitude diagram (Fig.\ \ref{figi}d)
confirms previous findings that
Hodge 11 is a very metal-poor cluster with a steep red giant
branch and a blue horizontal branch.

We confirm the finding of Walker (1993) that the distribution
of horizontal branch stars in Hodge 11 is skewed to the extent
that only blue HB stars appear.  We find 124 blue HB stars,
no RR Lyrae stars, and no red HB stars.
Graham \& Nemec (1984) and Walker (1993)
have previously surveyed Hodge 11 for RR Lyrae stars but none were
found.

We do not detect any significant gaps in the distribution of blue HB stars.
This contradicts the finding of Walker (1993)
that the blue HB stars in Hodge 11 are concentrated
into two groups
with a gap existing near $V\approx19.6$ mag.
Walker finds 43 blue HB stars whereas
we find 124 and it is quite probable that the gap
seen in Walker's Fig.\ 3 is simply due to counting statistics.

\section{THE AGE OF HODGE 11}

One of the most important goals of galactic astronomy is to understand
how and when galaxies form and evolve.
The oldest star clusters of the Magellanic Clouds
preserve important information about the formation and
evolution of these satellite galaxies of the Milky Way.
Although an extensive literature exists on the properties of the
Large and Small Magellanic Clouds,
we know surprisingly little about their oldest stellar populations.
This lack of knowledge about the formation epoch of the LMC and SMC,
has greatly hampered
our understanding of their physical and chemical evolution.

The ages of the oldest Large Magellanic Cloud star clusters
are virtually unknown (Da Costa 1993).
Besides Hodge 11, only five other LMC globular clusters
have published CCD-based color-magnitude diagrams
(NGC 1466: Walker 1992a;
NGC 1841: Walker 1990;
NGC 2210: Reid \& Freedman 1994;
NGC 2257: Walker 1989, Testa \et 1995;
Reticulum: Walker 1992b).
Most of these CMDs were produced while investigating the
RR Lyraes in these globular clusters.  As a result, while they
are all deep enough to determine the morphology of the horizontal
branch, only one (NGC 2257: Testa \et 1995) has photometry that
reaches below the main-sequence turnoff.

Faced with the lack of accurate main-sequence photometry in the LMC
globular clusters,
researchers have used the horizontal-branch morphologies
of these systems to study their ages relative to those of
the Galactic halo globular
clusters.
Lee (1992)
has shown that the HB morphology for fixed \feh\ varies with Galactocentric
distance.
Lee (1989) introduced the HB type index
$(B-R)/(B+V+R)$ where B, R, and V, respectively,
denote the number of blue HB stars, red HB stars, and RR Lyraes.
The more distant clusters tend to have redder HB types and
the scatter of HB type at a given \feh\ increases with Galactocentric distance.
This is indicative of the ``second parameter'' effect and is
consistent with the halo formation model of Searle \& Zinn (1978).
Lee (1992)
used his diagnostic \feh\ vs.\ $(B-R)/(B+V+R)$ diagram to
compare 7 LMC and SMC globular clusters with Galactic globular
clusters and found evidence that the LMC and SMC clusters formed
about 2 Gyr after the formation of the inner Galactic halo
globular clusters.
Using the same technique, Walker (1992a) and Da Costa (1993)
reached a similar conclusion that the age difference was 2--3 Gyr.

We find that the Lee HB morphology index
for Hodge 11 is $1.00^{+0.00}_{-0.04}$ with 124 blue HB stars,
0 RR Lyraes, and 0 red HB stars.  The lower limit is assumed to be
due to counting statistics alone.
The diagnostic \feh\ vs.\ $(B-R)/(B+V+R)$ diagram unfortunately
becomes degenerate with very blue (or very red)
horizontal branches, so Lee's HB index can not provide
a useful estimate of the relative age difference between Hodge 11
and the Galactic halo globular clusters.

The robust mean $\bmv$ color as a function of $V$ magnitude
of the Hodge 11 main-sequence and subgiant branch is
listed in Table \ref{tablethree}
and is shown graphically in Figs.\ \ref{figii} and \ref{figiii}.
This fiducial sequence was derived using 0.2 mag bins and the
robust mean color was determined after
3$\sigma$ outliers were iteratively rejected.
The reliability of this method can be checked in Fig.\ \ref{figii}
by comparing the robust mean \bmv\ color (filled circles) with
the median $\bmv$ color (open diamonds) of all the stars
in the same 0.2 mag bins.
The median and robust-mean \bmv\ colors agree within the errors
determined for the robust mean \bmv\ colors.

\placetable{tablethree}

\placefigure{figii}

\placefigure{figiii}

Spectroscopic abundance determinations by Cowley \& Hartwick (1982)
and Olszewski \et (1991) agree that Hodge 11 is metal-poor with
$\feh = -2.1\pm0.2$ dex.
The steep red giant branch revealed in the
color-magnitude diagrams of Stryker \et (1984), Walker (1993), and
our Figs.\ \ref{figi}\ and \ref{figiii} are in complete agreement
with the spectroscopic determination of the metallicity of Hodge 11.

We compare the Hodge 11 fiducial sequence with that of the metal-poor
Galactic globular cluster M92 (Stetson \& Harris 1988)
in Figs.\ \ref{figii}\ and \ref{figiii}.
We find that by making the M92 fiducial sequence fainter by
$\Delta V=4.05$ mag and adding an additional reddening of
$\Delta(\bmv)=0.055$ mag gives an excellent fit to the Hodge 11
fiducial sequence.
The upper and lower limits of these shifts,
as shown in Fig.\ \ref{figii},
are
$\Delta V=4.05\pm0.05$ mag
and
$\Delta(\bmv)=0.055\mp0.010$
mag.

Stetson \& Harris (1988) found that the apparent distance modulus of M92
is $(m-M)_V\approx14.6$ mag and adopted the reddening
of $\ebmv=0.02$ mag.
The reddening of Hodge 11 is thus
$
\ebmv_{\rm H11}
\approx
\ebmv_{\rm M92} + \Delta(\bmv)
\approx 0.075 \pm 0.005
$ mag which is in excellent agreement with Walker's (1993) estimate
of $\ebmv=0.08\pm0.02$ mag.
The apparent distance modulus of Hodge 11 is
$
(m-M)_{V,{\rm H11}}
\approx
(m-M)_{V,{\rm M92}}
+
\Delta V
\approx
18.65\pm0.12
$ mag where we have assumed an error of 0.10 mag for $(m-M)_{V,{\rm M92}}$.
The true distance modulus of Hodge 11 is then
$(m-M)_o\approx18.42\pm0.12$ mag assuming that
$A_V=3.1\ebmv\approx0.23\pm0.02$ mag
(Savage \& Mathis 1979).

Hodge 11
is $4\fdg71$ (Mateo 1987) from the rotation center of the LMC
given by Rohlfs \et (1984).
Walker (1993) notes that
if it is assumed that Hodge 11 lies in the LMC disk
then the inclination correction for the distance modulus is 0.09 mag
in the sense that Hodge 11 is closer to us than the LMC center.
If we add this inclination correction to our determination of the
distance modulus of Hodge 11 we then derive a distance modulus for
the LMC of $(m-M)_o=18.51\pm0.17$ mag which is in excellent agreement to
the value 18.5 mag of Westerlund (1990) and van den Bergh (1991).
For comparison, Crotts, Kunkel, \& Heathcote (1995) find the
distance modulus of the LMC to be $18.57\pm0.13$ mag based on light travel
time measurements across the ring of SN 1987A.

We can determine the relative age difference between M92 and Hodge 11 in the
following manner.
We estimate the absolute visual magnitude
of the main-sequence turnoff of Hodge 11
to be
$V_{\rm TO,H11}\approx22.65\pm0.10$ mag.
The absolute visual magnitude of the main-sequence turnoff of Hodge 11 is
then
$
M_{V,{\rm H11}}^{\rm TO}
\approx
4.00\pm0.16
$
mag.
The absolute visual magnitude of the main-sequence turnoff of M92 is
$
M_{V,{\rm M92}}^{\rm TO}
\approx
4.00\pm0.14
$
mag
assuming $V_{\rm TO,M92} = 18.60\pm0.10$ mag
(Stetson \& Harris 1988).
The {\em difference} between these turnoffs is
$
M_{V,{\rm H11}}^{\rm TO}
-
M_{V,{\rm M92}}^{\rm TO}
=
0.00\pm0.21
$ mag
which gives an age resolution of
$\sim$21\%
using Eq.\ 2 of Mighell \& Butcher (1992).
We have thus found that
$
age_{\rm H11}
/
age_{\rm M92}
\approx
1.00\pm0.21
$
which translates to
$
age_{\rm H11}
\approx
15\pm3
$
Gyr if M92 is assumed to be 15 Gyr old.
This is probably a conservative estimate of the true relative age-difference
uncertainty between M92 and Hodge 11.  If we naively
assume that the only source of uncertainty is our error in determining
$V_{\rm TO,H11}$ and the $V$ magnitude
shift between the M92 and Hodge 11 fiducial sequences
then we find that
$
M_{V,{\rm H11}}^{\rm TO}
-
M_{V,{\rm M92}}^{\rm TO}
=
0.00\pm0.11
$ mag
which gives an optimistic age resolution of
$\sim$10\%
using Eq.\ 2 of Mighell \& Butcher (1992).
The relative age-difference uncertainty between Hodge 11 and M92
is probably between 10 and 21 percent.

Our analysis suggests that, within the accuracy of our photometry,
Hodge 11 is as old as M92 --- probably one of the oldest globular
clusters in the Milky Way.
Furthermore, if we assume
that the globular cluster Hodge 11 has always been a part
of the Large Magellanic Cloud and was
not stripped from the halo of the Milky Way or
absorbed from a cannibalized dwarf spheroidal galaxy,
then the oldest stars in the
Large Magellanic Clouds and the Milky Way appear to have the same age.

\bigskip
\bigskip
Support for this work was provided by NASA through grant
Nos.\ STSCI GO-5386, GO-5464, and GO-5475
from the Space Telescope Science Institute,
which is operated by the Association of
Universities for Research in Astronomy, Inc.\ under
National Aeronautics and Space Administration (NASA)
contract NAS5-26555, and by grant No.\ NAGW-2479 from the
Long Term Space Astrophysics Research Program.

\newpage
\centerline{FIGURE CAPTIONS}

\figcaption{\label{figv}
Comparison of our photometry with Walker (1993).
We have 80 stars in common on the WF4 CCD of the WFPC2 instrument.
A robust estimate of the mean difference in $V$
(using $2$$\sigma$ rejection of outliers)
is $0.009\pm0.010$ mag
with our $V$ magnitudes being $\sim$1\% fainter.
This $V$ zeropoint difference is not statistically significant.
The seeing varied from 1.5 to 2.0 arcsec when
Walker (1993) observed Hodge 11.
Almost all of the stars with large $V$ magnitude differences
contain more than one star within a circular aperture of
radius 1.0 arcsec (10 pixels) on the WF4 CCD.
}

\figcaption{\label{figi}
The $V$ \vs $\bmv$ color-magnitude diagram of the observed stellar
field in the LMC globular cluster Hodge 11.
The {\em Hubble Space Telescope} WFPC2 instrument
was used to make
one 300-s observation with the F555W filter
and
one 600-s observation with the F450W filter.
(a)
The 13175 stars are plotted with filled circles
and the 1855 CCD defects are plotted with open circles.
The CCD defects were identified using the procedure described in the text.
(b) The 12281 stars within 83.2\arcsec of the center of the globular
cluster are plotted.
(c) The 894 stars beyond 83.2\arcsec of the center of the globular
cluster are plotted.
(d) The ``cleaned'' color-magnitude diagram of Hodge 11 with 9506 stars.
The details of the statistical field-subtraction are described in the
text.
}

\figcaption{\label{figii}
The fiducial sequence of Hodge 11 (filled circles)
compared with the M92 fiducial sequence of Stetson \& Harris (1988).
The M92 fiducial sequence is shown from left to right assuming
a shift in $\bmv$ color of 0.045, 0.055, and 0.065 mag
and a shift in $V$ magnitude of 4.10, 4.05, and 4.00 mag, respectively.
The fiducial sequence of Hodge 11 was derived using 0.2 mag bins and the
robust mean color was determined after $3\sigma$
outliers were iteratively rejected.
The reliability of this
method is checked by comparing the robust mean \bmv\ color (filled circles)
with the median \bmv\ color (open diamonds) of all the stars in the
same 0.2 mag bins.
}

\figcaption{\label{figiii}
The ``cleaned'' $V$ \vs $\bmv$ color-magnitude diagram
of LMC globular cluster Hodge 11 compared with M92 fiducial sequences.
The Hodge 11 fiducial sequence is plotted with open circles.
The lower curve is the M92 fiducial sequence of
Stetson \& Harris (1988).
The solid curves on the left and at the top are,
respectively,
the fiducial sequences
of the M92 horizontal branch and the M92 asymptotic giant branch of
Buonanno, Corsi, \& Fusi Pecci (1985).
All M92 fiducial sequences are shown
assuming $\Delta(\bmv)=0.055$ mag and $\Delta V=4.05$ mag.
}

\begin{table}
\dummytable\label{tableone}
\end{table}

\begin{table}
\dummytable\label{tabletwo}
\end{table}

\begin{table}
\dummytable\label{tablethree}
\end{table}

\end{document}